# Fair Exchange of Digital Signatures using RSA-based CEMBS and Offline STTP

Jamal A. Hussein, Mumtaz A. AlMukhtar

**Abstract**—One of the essential security services needed to safeguard online transactions is fair exchange. In fair exchange protocols two parties can exchange their signatures in a fair manner, so that either each party gain the other's signature or no one obtain anything useful. This paper examines security solutions for achieving fair exchange. It proposes new security protocols based on the "Certified Encrypted Message Being Signature" (CEMBS) by using RSA signature scheme. This protocol relies on the help of an "off-line Semi-Trusted Third Party" (STTP) to achieve fairness. They provide with confidential protection from the STTP for the exchanged items by limiting the role and power of the STTP. Three different protocols have been proposed. In the first protocol, the two main parties exchange their signatures on a common message. In the second protocol, the signatures are exchanged on two different messages. While in the third one, the exchange is between confidential data and signature.

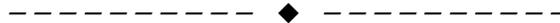

## 1 INTRODUCTION

THE third party is necessary in fair exchange protocols, the fairness can not be guaranteed without help of the third party. The third party may be online, i.e. involved in each transaction to help each party to gain the others signature, or it may be offline, i.e. it involved only when an problem occur while the two main parties try to exchange their signatures, in this case the third part help to recovering the signatures. Furthermore, the third party may be full trusted (TTP) or semi trusted (STTP), trusted means that the third party can obtain the main parties' signatures while trying to recover the signatures, while semi-trusted means that the third party can only help to recover the signature without revealing the signature to the third party because the third party may misbehave by itself.

## 2 RELATED WORKS

There are many protocols that based on CEMBS found in the literature. The CEMBS is first introduced in [1] by Feng Bao, Robert H. Deng and Wenbo Mao. In [1] two types of CEMBS is proposed, the first based on GQ signature scheme and the second based on DSA signature scheme, each one of these CEMBS certificates are based on offline TTP. In [2] the CEMBS is based on GQ with offline STTP by using blind decryption. Blind decryption means that the STTP partially decrypt the ciphertext where trying to recover the signature. GQ-based CEMBS has one severe flaw, for in this scheme only one component D of signature (d, D) is encrypted. And it is an observation that the verifier can recover D by the virtue of the publicity of d, and by calculation of the inverse of mod q, not mod n. It is just this flaw that is exploited in

[6] to successfully break the fair exchange protocol based on this kind of CEMBS [3].

In DSA-based CEMBS have the following drawbacks: (1) CEMBS construction and verification involve a lot of calculations (that is why DSA is slower than RSA). (2) The structure of resulting CEMBS certificate is too complex and lengthy. (3) The DSA is the standard used by (NSA) so it afraid to have a trap door.

The CEMBS-based on RSA is first presented in [3]; the first disadvantage of this type of CEMBS is that it based on online TTP, which means that TTP is involved in each transaction. Finding such trusted third party is not easy and involving third party in each transaction may cause bottleneck. Furthermore the third party may misbehave by itself. The second disadvantage is there are more calculations in the constructing of and verifying of CEMBS certificate.

In this paper a new RSA-based CEMBS is presented base on offline STTP. The third party is only involved when one of the main parties misbehaves or an error occurs in communication channel, this decrease the load on the third party. Also the third party is semi trusted, that is he cannot obtain the main parties' signatures, he only help them to recover the signature when an error occurs. These features are implemented by using blind decryption. This RSA-based CEMBS is more efficient in computation and communication comparing with other signature schemes that used in constructing CEMBS certificate.

## 3 FAIR EXCHANGE PROBLEM

Fair exchange of digital signatures or documents between two distrusted parties (e.g. individuals or companies) is one of the major issues in e-commerce systems. A fair exchange protocol allows two potentially mistrusted parities to exchange their digital signatures over the Internet in a fair way, so that either each of them obtains the other's signature, or neither party does.

---

• J.A.Hussein is with the Computer Department, College of Science, University of Sulaimani. Sulaimani, Kurdistan-Iraq.
• M. A. AlMukhtar, Baghdad, Iraq.

## 4 REQUIRMENTS

**Client A**: one of the main parties involved in the protocol. She is responsible for initiating the executing of the protocol.
**Client B**: the second main party. He is responsible for executing the recovery sub-protocol when an error occurs.
**STTP**: the semi-trusted third party, help the main parties to recover the signatures when one of the main parties is misbehave or an error occurs in the communication channel.

## 5 NOTATIONS

| | |
|---|---|
| $SK_T, PK_T$ | Elgamal cryptography scheme private/public keys for STTP |
| $SK_i, PK_i$ | Elgamal cryptography scheme private/public keys for Client i |
| $d_i, e_i$ | Client i's private/public key (Signature scheme) |
| $M_i$ | Client i's message |
| $S_i$ | RSA signature on a message |
| $W_i, V_i$ | Ciphertext of Elgamal cryptography scheme |
| $\mathcal{C}$ | Blind ciphertext of Client A |
| $c_i, r_i$ | CEMBS certificate of party i |
| $H(.)$ | One-way hash function |
| $N_i$ | Modulus of RSA signature scheme for client i |
| $G$ | Generator of nA |
| $P_i, G_i$ | Elgamal parameters for Party i |
| $\lvert x \rvert$ | Size of x in bits |
| $X \mid\mid Y$ | Concatenation of X and Y |

## 6 RSA SIGNATURE SCHEME

RSA is proposed cryptosystem in [4] by Rivest, Shamir and Adleman. We can use this system as a signature scheme.

**RSA key generation**. The signer chooses two large secret primes p and q, and calculates public modulus n as n = pq, selects number e, such that $0 < e < \varphi(n)$ and e is relatively prime to $\varphi(n)$. Function $\varphi(.)$ is Euler's totient function. There exists an inverse d of e modulo $\varphi(n)$, i.e. $d = e^{-1} \mod \varphi(n)$. The party's public key is e and its private key d.

**RSA signing**. $s = m^d \mod n$
Where s is the signature on m, m is the massage to be signed.
**RSA verification**. To verify that s is really the signer's signature on m, we verify if
$m = s^e \mod n$ = yes or no
If the result is yes then s is the signer's signature on m.

## 7 ELGAMAL CRYPTOGRAPHY SCHEME

The Elgamal cryptosystem proposed in [5] is the cryptographic scheme that used by CEMBS primitives.

**Elgamal Key Generation**. To generate the Elgamal system keys, first, a suitable prime P is chosen such that the discrete logarithm problem is difficult for integers less than P. The suitable PK, g, and SK are chosen where g is the gernerator of P and $PK = g^{SK} \mod P$. PK, g and P are then made public and SK kept private.

**Elgamal Encryption Process**. To encipher the plaintext m, a secret random integer w is choosing such that $w < (P - 1)$, the ciphertext is (W, V)
Where $W = G^w \mod P$ and $V = m \, PK^w \mod P$.

**Elgamal Decryption Process**. the decryption process is:
$m = C \, (W^{SK})^{-1} \mod P$

## 8 BLIND DECRYPTION

Blind decryption for Elgamal system can be done as follows. The ciphertext receiver (who has no SK) gives the decryptor (who has SK) only W while keeps V to himself. The decryptor computes $W^{SK}$ and sends back to the receiver. As a result, only the receiver but not the decryptor can obtain m.
In our fair exchange protocols with offline STTP, B is the ciphertext receiver and STTP is the decryptor.

## 9 SYSTEM INIZIALIZAITON

**Client A**. client A chooses two prime numbers p, q so that $\lvert p \rvert = \lvert q \rvert = 512$, sets $n_A = pq$ and chooses $g \in Z_n^*$ to the generator of n. $e_A \in Z_n^*$ relatively prime to $\varphi(n)$. Sets $d_A = e_A^{-1} \mod \varphi(n)$. $e_A$ is the public key for RSA signature scheme and d is the private key for RSA signature scheme.
A chooses $P_A$ as a prime number so that $\lvert P_A \rvert = 1024$. Let $G_A$ be the generator of $P_A$, chooses $SK_A \in Z_{P_A}$ and sets $PK_A = G_A^{SK_A} \mod P_A$. $(PK_A, SK_A)$ is the public/private key pair.
**Client B**. client B chooses two prime numbers p, q so that $\lvert p \rvert = \lvert q \rvert = 512$, sets $n_B = pq$. $e_B \in Z_n^*$ relatively prime to $\varphi(n)$. Sets $d_B = e_B^{-1} \mod \varphi(n)$. $e_B$ is the public key for RSA signature scheme and d is the private key for RSA signature scheme.
**STTP**. chooses the prime $P_T$ so that $\lvert P_T \rvert = 1024$, let $G_T$ be the generator of $P_T$ (order of $G_T$ is large). $SK_T \in Z_{P_T}$ sets $PK_T = G_T^{SK_T} \mod P_T$.

## 10 CEMBS DEFINITION

Let s be the signature on the public message m under d and (W, V) be the ciphertext of the signature s under PK.
Let (r, c) be the CEMBS certificate, there exist a public verification algorithm. The receiver generates a blind ciphertext and then implements the verification algorithm.

If the result is yes, then (W, V) is the ciphertext s under PK and s is the signature on m under d.

The STTP can help the receiver to recover the signature from the CEMBS certificate by blindly decrypt the blind ciphertext without revealing the signature to STTP.

## 11 CEMBS GENERATION AND VERIFICATION

**Signing**: Assume that m is the public message. s = $m^d$ mod n is the signature on the message m.

**Encryption**: choose $w \in_R Z_P^*$, the ciphertext is (W, V), where W = $G^w$ mod P and V = s (PK) mod P.

**CEMBS Generation**: choose $u \in_R Z_n^*$, |u| = 400
c = H (g || W || $\mathcal{C}$ || a || A)
Where C = $g^V$, a = $G^u$, A = $(G^{PK})^u$
r = u – cw
the CEMBS certificate is r and c

**CEMBS verification**: we check whether
c ?= H(g || W || $\mathcal{C}$ || $G^r W^c$ || $(G^{PK})^r (W^{PK})^c$)

## 12 PROOF OF CORRECTNESS

The correctness of RSA-based CEMBS is ensured by proving the correctness of the CEMBS generation and verification.

We have

$g^u = g^{(r+cw)} = g^r g^{cw} = g^r (g^w)^c = g^r W^c$

Also we have

$(g^{PK})^u = (g^{PK})^{(r+cw)} = (g^{PK})^r (g^{PK})^{cw} = (g^{PK})^r ((g^{PK})^w)^c$

$= (g^{PK})^r ((g^w)^{PK})^c = (g^{PK})^r (W^{PK})^c$

## 13 PROTOCOL 1: FAIR EXCHANGE OF SIGNATURES ON A COMMON MESSAGE

This section explains the implementation of the proposed CEMBS on a common message m.

**Basic sub-protocol steps**

1. A computes her signature $s_A$, encrypts $s_A$ under STTP's public key $P_{KT}$ to generate $W_A$ and $V_A$, and generate the CEMBS certificate (r, c)

$s_A = m^{d_A}$ mod $n_A$

$W_A = G_T^w$ mod $P_T$     where w < (P - 1)

$V_A = s_A (P_{KT})_w$ mod $P_T$

$c_A$ = H (g || $W_A$ || C || a || A)     Where $u \in_R Z_n^*$, |u| = 400
C = $g^V$, a = $G_T^u$, A = $(G_T^{P_{KT}})^u$

$r_A = u - c_A w$

A → B: $W_A$, $V_A$, $c_A$, $r_A$

2. B, upon receiving ($W_A$, $V_A$, $c_A$, $r_A$), sets blind ciphertext $\mathcal{C}$, and checks whether
$c_A$ ?= H(g || $W_A$ || $\mathcal{C}$ || $G_T^r W_A^{c_A}$ | $(G_T^{PK_T})^{r_A} (W_A^{PK_T})^{c_A}$)
if the answer is 'no' then B stops the protocol. If it is 'yes', B computes his signature $s_B = m^{d_B}$ mod $n_B$ and send it to A

B → A: $s_B$

3. A, after receiving $s_B$, checks whether $s_B$ ?= $m^{e_B}$ mod $n_B$, if it is 'no', A stops the protocol; if it is 'yes' A sends his signature to B

A → B: $s_A$

4. B, after receiving $s_A$, checks whether $s_A$ ?= $m^{e_A}$ mod $n_A$. If it is valid, B accepts the signature and the protocol is ended successfully.

**Recovery sub-protocol steps**

1. If B does not receive any thing or the received $s_A$ is invalid, he sets the blind ciphertext $\mathcal{C}$ for A's ciphertext and encrypts his signature under A's public key $e_A$, generates his CEMBS, and send each of A's CEMBS certificate, A's blind ciphertext, his ciphertext and CEMBS to STTP.

$\mathcal{C} = g^V$

$W_B = G_A^w$ mod $P_A$     where w < (P - 1)

$V_B = s_B (PK_A)^w$ mod $P_A$

$c_B$ = H (g || $W_B$ || $V_B$ || a || A)     Where $u \in_R Z_n^*$, |u| = 400
a = $G_A^u$, A = $(G_A^{PK_A})^u$

$r_B = u - c_B w$

B → STTP: $W_A$, C, $c_A$, $r_A$, $W_B$, $V_B$, $c_B$, $r_B$

2. STTP, upon receiving the two ciphertexts and CEMBS certificates, verify the two CEMBS,
$c_A$ ?= H(g || $W^A$ || $\mathcal{C}$ || $G_T^r W_A^{c_A}$ || $(GT^{PK_T})r_A (W_A^{PK_T})^{c_A}$)
$c_B$ ?= H(g || $W_B$ || $V_B$ || $G_A^{r_B} W_B^{c_B}$ || $(G_A^{PK_A})^{r_B} (W_A^{PK_A})^{c_B}$)
if the verification is OK then he blindly decrypt the blind ciphertext of A and send the blind decryption of A's signature to B, and B's ciphertext to A

STTP → B: $(W_A)^{SK_T}$
STTP → A: $W_B$, $V_B$

3. B, receives the blind ciphertext of A and perform the remaining decryption to recover $s_A$, and then check the $s_A$ validity.
A receives the B's ciphertext, decrypt it to obtain $s_B$.

## 14 PROTOCOL 2: FAIR EXCHANGE OF SIGNATURE ON DIFFERENT FILES

Here it is assumed that A and B have agreed on two files $M_A$ and $M_B$. Client A sign the $m_A$ where $m_A = M_A$ || $H(M_B)$ and

Client B sign $m_B$ where $m_B = M_B || H(M_A)$. The steps of the basic and recovery subprotocols are same as Protocol 1.

## 15 PROTOCOL 3: FAIR EXCHANGE OF CONFIDENTIAL DATA AND SIGNATURE

This protocol is used to exchange confidential data and a signature on the message between A and B. The protocol lets B send a message M to A in the exchange for A's signature on H(M).

**Basic sub-protocol steps**

1. Party A computes her signature $s_A = (h(M))^{d_A} \mod n_A$ and the ciphertext $W_A$ and $V_A$. A then generates CEMBS $(r_A, c_A)$. A sends
A → B: $W_A, V_A, c_A, r_A$
2. B, receive $(W_A, V_A, c_A, r=)$, converts $V_A$ to blind ciphertext $\mathcal{C}$ checks whether the CEMBS certificate is valid. If it is valid, B sends
B → A : M
otherwise, B stops the protocol.
3. After receiving M, A checks whether it matches h(M). If yes, A sends
A → B : $s_A$
Otherwise, A does nothing.
4. B receives $s_A$ and checks its validity. If it is valid, B accepts the signature and the protocol terminates.

**Recovery sub-protocol steps**

1. If B does not receive or receive an incorrect $s_A$, he converts the ciphertext $V_A$ to a blind ciphertext $\mathcal{C}$, encrypts M under A's public key, creates CEMBS and then send the two ciphertext and the two CEMBS to STTP.
$\mathcal{C} = g^V$

$W_B = G_A^w \mod P_A$      where $w < (P - 1)$

$V_B = M (PK_A)^w \mod P_A$

$c_B = H (g || W_B || V_B || a || A)$    Where $u \in_R Z_n^*$, $|u| = 400$, $a = G_A^u, A = (G_A^{PK_A})^u$

$r_B = u - c_B w$

B → STTP: $W_A, \mathcal{C}, c_A, r_A, W_B, V_B, c_B, r_B$

2. STTP, upon receiving the two ciphertexts and CEMBS certificates, verify the two CEMBS,
$c_A ?= H(g || W_A || C || G_T^{r_A} W_A^{c_A} || (G_T^{PK_T})^{r_A} (W_A^{PK_T})^{c_A})$
$c_B ?= H(g || W_B || V_B || G_A^{r_B} W_B^{c_B} || (G_A^{PK_A})^{r_B} (WA^{PK_A})^{c_B})$
if the verification is OK then he blindly decrypt the blind ciphertext of A and send the blind decryption of A's signature to B, and B's ciphertext to A

STTP → B: $(W_A)^{SK_T}$
STTP → A: $W_B, V_B$
3. B, receives the blind ciphertext of A and perform the remaining decryption to recover $s_A$, and then check the $s_A$ validity.
4. A receives the B's ciphertext, decrypt it to obtain M.

## 16 SECURITY OF THE PROTOCOL

It is easy to see that A and B obtains each other's signatures without any involvement of STTP.
B has two chances to perform improperly. The first one is where B may send A an incorrect $s_B$, but A can detect this and refuse to give $s_A$ to B. The second chance is right after receiving ciphertext and CEMBS certificate of A, B stops the protocol, goes to STTP, and asks it to decrypt $W_A$ in order to get $s_A$ while without giving $s_B$ to A, however STTP will open $W_A$ for B only if B gives correct $W_B, V_B$ to STTP and STTP will forward it to A.
A may perform improperly by giving B incorrect ciphertext and CEMBS certificate. If A performs improperly later by sending B an incorrect $s_A$ or not sending anything, B can ask STTP to decrypt $W_A$ and get A's signature. Note that STTP also sends B's ciphertext to A in this case.

## 18 CONCLUSION

In this paper a new RSA-based CEMBS is proposed. This CEMBS is constructed by using RSA signature scheme and Elgamal Cryptography scheme. It is used to convince the receiver that the encrypted message is really the sender's signature while without revealing the signature to the receiver. The blind decryption is used to ensure that the third party can only help the client B to recover the client A's signature without disclosing the signature to the third party. The third party is offline, i.e. only involved in the protocol when a problem occurs.
A protocol is presented that enables two parties to exchange their signature on a common message. We can easily modify the protocol to exchange signature on a two different messages. To exchange on different messages, each client signs his/her own message on a different message, create CEMBS certificate, and send the signatures, the messages, and CEMBS certificates.